\documentclass[twocolumn,showpacs,preprintnumbers,amsmath,amssymb]{revtex4}
\usepackage[dvips]{graphicx}
\begin{document}
\title{Relativistic diffusion process and analysis of transverse momentum distributions observed at RHIC}
\author{Naomichi Suzuki}
 \affiliation{Department of Comprehensive management, Matsumoto University, Matsumoto, Japan}
 \email{suzuki@matsu.ac.jp} 
\author{Minoru Biyajima}
 \affiliation{Department of Physics, Shinshu University, Matsumoto, Japan }
 \email{mbiyajima@azusa.shinshu-u.ac.jp}
\begin{abstract}
Large transverse momentum distributions of identified particles observed at RHIC are analyzed by a  relativistic stochastic model in the three dimensional rapidity space. Temperature for inclusive reactions is estimated.
\end{abstract}
\pacs{25.75.-q, 13.85.Ni, 02.50.Ey}

\maketitle
\section{Introduction}
   
At RHIC colliding energy of nuclei becomes up to 200 GeV, and thousands of particles 
are produced per event.  To describe such many particle system, 
a sort of collective approach will be useful. 
One-particle rapidity or pseudo-rapidity distributions observed at RHIC 
are well described by the Ornstein-Uhlenbeck process~\cite{biya02}.

As for the transverse momentum $p_T$ distribution observed at RHIC, it has a long tail in the GeV region compared with an exponential distribution in $p_T$.

In reference~\cite{minh73}, an empirical formula for large $p_T$ distributions at polar angle $\theta=\pi/2$, 
\begin{eqnarray}
  E\frac{d^3\sigma}{d^3p}\left|_{\theta=\pi/2} \right. &=& A\exp[-y_T^2/(2L_T)],
    \nonumber \\
  y_T &=& \frac{1}{2}\ln \frac{E+|{\bf p}_T|}{E-|{\bf p}_T|},    \label{eq.int1}
\end{eqnarray}
was proposed from the analogy of Landau's hydrodynamical model. In Eq.(\ref{eq.int1}), $E$ denotes energy of an observed particle, $L_T$ is a parameter, and $y_T$ is called the transverse rapidity.  Equation (\ref{eq.int1}) well describes the large $p_T$ distributions for $p+p\rightarrow \pi^0 + X$ and $p+p\rightarrow \pi^\pm + X$. However, it cannot be derived from the hydrodynamical model.  

The transverse rapidity is defined in the geodesic cylindrical coordinate system in the three dimensional  rapidity space .  
The Lorentz invariant phase volume element in it is given as
 \begin{eqnarray*}
  \frac{d^3p}{E}=m^2\sinh\xi\,\cosh\xi\, dy d\xi d\phi.
 \end{eqnarray*}
In the above equation, $y$ denotes the longitudinal rapidity, and $\xi$ denotes the transverse rapidity;
 \begin{eqnarray*}
   y=\ln\frac{E+p_L}{m_T}, \quad \xi=\ln\frac{m_T+|{\bf p}_T|}{m},
 \end{eqnarray*}
where $E, p_L$, ${\bf p}_T$ and $m$ denote energy, longitudinal momentum, transverse momentum, and mass of the observed particle, respectively, and $m_T=\sqrt{{\bf p}_T^2 + m^2}$.
It should be noted that $y_T$ coincides with $\xi$, only if $\theta=\pi/2$.

  We have proposed the relativistic diffusion model, and analyzed large $p_T$ distributions for charged particles in $Au+Au$ collisions~\cite{suzu04}. 
  The distribution function of it is gaussian-like in radial rapidity, and resemble with Eq.(\ref{eq.int1}) at $\theta=\pi/2$.
  
  In section 2, the relativistic diffusion model is briefly explained. In section 3, analyses of large $p_T$ distributions for identified particles, $\pi^0$, $\pi^-$, $K^-$ and $\bar{p}$ observed at RHIC~\cite{adle03a,adle03b,adle03c} are made.  Temperature is also estimated from $p_T$ distributions. Final section is devoted to summary and discussions.
\section{Diffusion equation in the three dimensional rapidity space}
For simplicity, we consider the diffusion equation with radial symmetry in the geodesic polar coordinate system,
\begin{eqnarray}
  \frac{\partial f}{\partial t}= \frac{D}{\sinh^2\!\rho}\, 
      \frac{\partial}{\partial \rho}\left( 
        \sinh^2\!\rho\, \frac{\partial f}{\partial \rho} 
      \right) .          \label{eq.dif1}
\end{eqnarray}
with initial condition
\begin{eqnarray}
    f(\rho,t=0)= \frac{\delta(\rho)}
     {4\pi\sinh^2\!\rho}.     \label{eq.dif2}
\end{eqnarray}
In Eq.(\ref{eq.dif1}), $D$ is a diffusion constant, and $\rho$ denotes the radial rapidity, which is written with energy $E$, momentum ${\bf p}$ and mass $m$ of observed particle,  
\begin{eqnarray}
   \rho=\ln \frac{E+|{\bf p}|}{m}.    \label{eq.dif3}
\end{eqnarray}
Inversely, energy and momentum are written respectively as
\begin{eqnarray}
    E = m\cosh\rho,   \qquad
       |{\bf p}| = \sqrt{p_L^2+ {\bf p}_T^2} = m\sinh\rho.   \label{eq.dif4}
\end{eqnarray}

The solution~\cite{karp59,suzu04b} of  Eq.(\ref{eq.dif1}) with the initial condition (\ref{eq.dif2}) is given by
\begin{eqnarray*}
  f(\rho,t) &=& \left( 4\pi Dt \right)^{-3/2} {\rm e}^{-Dt}
     \frac{\rho}{\sinh\rho}
     \exp \left[ -\frac{\rho^2}{4Dt} \right].   
 \end{eqnarray*}

A physical picture described by Eqs.(\ref{eq.dif1}) and (\ref{eq.dif2}) is as follows;
After a collision of nuclei, particles are produced at the origin of rapidity space expressed by Eq.(\ref{eq.dif2}). Then those particles propagate according to the diffusion equation (\ref{eq.dif1}).  In the course of the space time development, energy is supplied from the leading particle system (collided nuclei) to the produced  particle system. Number density of particles becomes lower and at some (critical) density, interactions among secondary particles cease, and particles become free.

We can analyze transverse momentum (rapidity) distributions at fixed polar angle $\theta$, using the equation,
\begin{eqnarray}
  f(\rho,t)&=& C \frac{\rho}{\sinh\rho} 
     \exp\left[-\frac{\rho^2}{2\sigma(t)^2} \right], \nonumber \\
   \sigma(t)^2 &=& 2Dt,    \label{eq.dif5}
\end{eqnarray}
with parameters, $C$ and $\sigma(t)^2$, 
\footnote{Somewhat different analyses based on statistical models and the different stochastic model have been made in M.Biyajima et al, hep-ph/0403063}.
where transverse momentum is given by $|{\bf p}_T|=m\sinh\rho\sin\theta$
%
%

%
\section{Analysis of $p_T$ distributions observed at RHIC}

\begin{figure}[htbp]
    \includegraphics[scale=0.50,bb=40 220 490 650,clip]{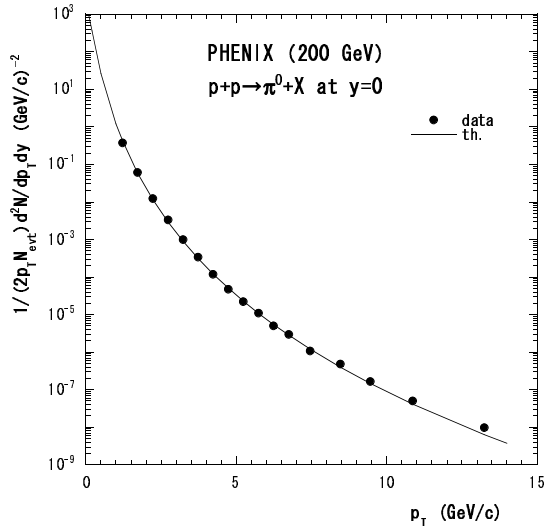} 
    \caption{\label{fig.pppi0}$p_T$ distribution for $p+p\rightarrow \pi^0+X$ at $y=0$ ~\cite{adle03a} }
 \end{figure}

\begin{figure}[htbp]
    \includegraphics[scale=0.45,bb=40 160 520 650,clip]{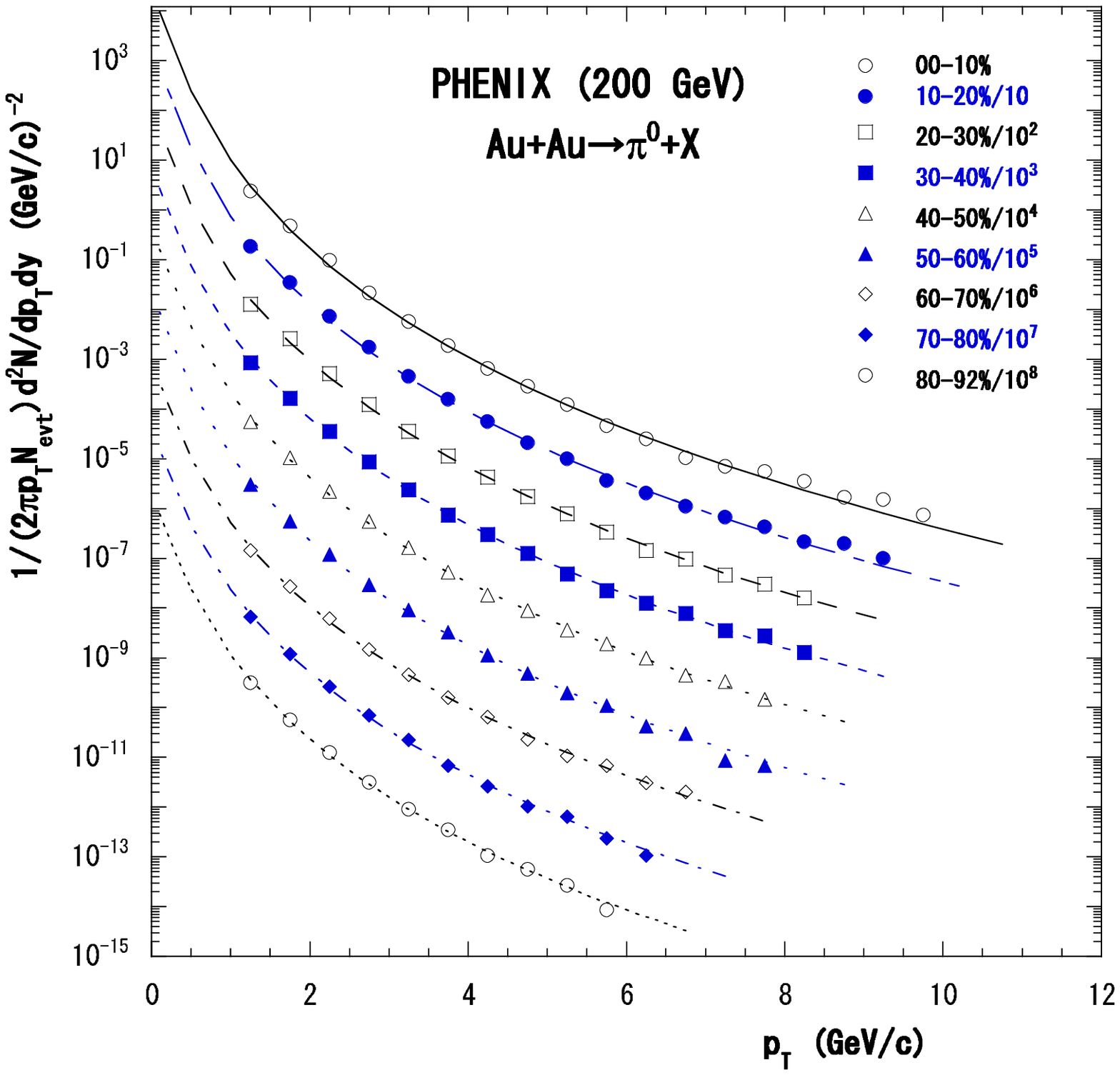}
    \caption{\label{fig.aapi0}$p_T$ distribution for 
      $Au+Au\rightarrow \pi^0+X$ at $y=0$~\cite{adle03b} }
 \end{figure}

Transverse momentum ($p_T$) distributions of identified particles observed by the PHENIX collaboration~\cite{adle03a,adle03b,adle03c} are analyzed.   
The results on $p_T$ distribution in $p+p \rightarrow \pi^0 +X$ is shown in Fig.~\ref{fig.pppi0}. Solid curve shown in Fig.1 is drawn by the use of Eq.(\ref{eq.dif5}), parameters of which are estimated with the least mean square method, and are shown in Table 1.
The results on $p_T$ distributions in $Au+Au\rightarrow\pi^0+X$ are shown in Fig.~\ref{fig.aapi0} and Table 2. 
Observed $p_T$ distributions on $\pi^0$ both in $p+p$ and $Au+Au$ collisions are well described by Eq.(\ref{eq.dif5}).
\begin{table}
 \caption{\label{tab.table1} Estimated parameters on $p_T$ distributions in $p+p \rightarrow \pi^0+X$    at $y=0$ at $\sqrt{s}=200$ GeV~\cite{adle03a}}
 \begin{ruledtabular}
 \begin{tabular}{cccc}  
     $C$                 & $\sigma(t)^2$    & $\chi^2_{min}$/n.d.f. \\ \hline
    1388.0 $\pm$ 121.5  & 0.602$\pm$ 0.004 & 12.4/15      \\ 
 \end{tabular}
 \end{ruledtabular}
 \end{table}
%
%
\begin{table}
 \caption{\label{tab.table2} Estimated parameters on $p_T$ distributions 
   in $Au+Au\rightarrow \pi^0+X$ at $y=0$ at $\sqrt{s}=200$ GeV~\cite{adle03b}}
 \begin{ruledtabular}
 \begin{tabular}{cccc} 
  centrality &  $C$  & $\sigma(t)^2$    & $\chi^2_{min}/n.d.f $ \\ \hline
   00-10\%  &  16088  $\pm$ 2003   & 0.574 $\pm$ 0.005 & 27.3/16 \\ 
   10-20\%  &  1161.5 $\pm$ 1390.1 & 0.580 $\pm$ 0.005 & 20.0/15 \\ 
   20-30\%  &  7345.9 $\pm$  922.7 & 0.586 $\pm$ 0.005 & 14.9/13 \\ 
   30-40\%  &  4378.8 $\pm$  569.3 & 0.593 $\pm$ 0.005 & 15.7/13 \\ 
   40-50\%  &  2559.8 $\pm$  345.1 & 0.600 $\pm$ 0.006 &  9.4/12 \\ 
   50-60\%  &  1425.3 $\pm$  194.3 & 0.599 $\pm$ 0.007 & 12.4/12 \\ 
   60-70\%  &   527.1 $\pm$   78.9 & 0.617 $\pm$ 0.008 &  8.2/10 \\ 
   70-80\%  &   241.2 $\pm$   39.6 & 0.616 $\pm$ 0.009 &  7.2/9  \\ 
   80-92\%  &   123.0 $\pm$   24.7 & 0.611 $\pm$ 0.011 &  5.0/8  \\ 
 \end{tabular}
 \end{ruledtabular}
 \end{table}
%

\begin{figure}[htbp]
    \includegraphics[scale=0.45,bb=40 160 520 650,clip]{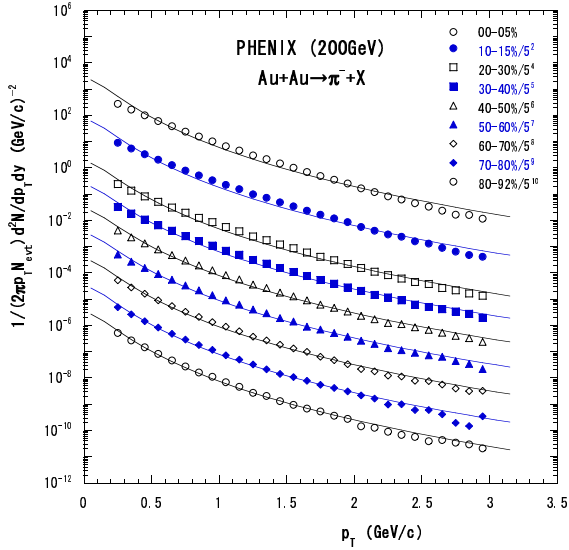} 
    \caption{\label{fig.aapin}$p_T$ distribution for $Au+Au\rightarrow \pi^-+X$ at $y=0$~\cite{adle03c} }
 \end{figure}

The results on $Au+Au\rightarrow\pi^-+X$ are shown in Fig.~\ref{fig.aapin} and Table 3. As is seen from Fig.3 and Table 3, fitting of the theoretical curve to the observed $p_T$ distribution is gradually improved as the centrality cut increases, in other words, the average number of participant nucleons decreases.

\begin{figure}[htbp]
    \includegraphics[scale=0.45,bb=40 150 540 650,clip]{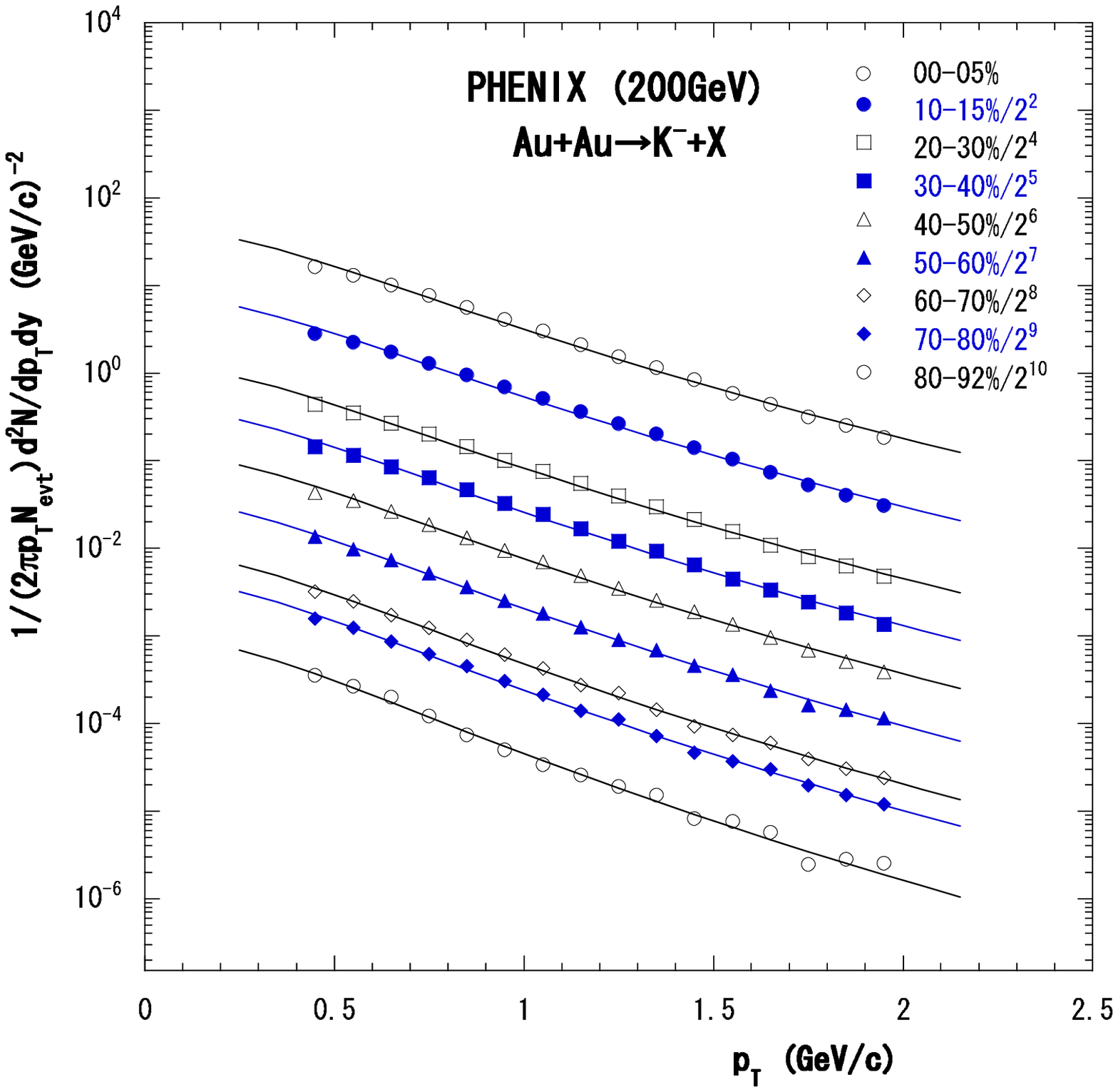} 
    \caption{\label{fig.aakn}$p_T$ distribution
        for $Au+Au\rightarrow K^-+X$ at $y=0$~\cite{adle03c} }
 \end{figure}

\begin{figure}[htbp]
    \includegraphics[scale=0.45,bb=40 160 540 650,clip]{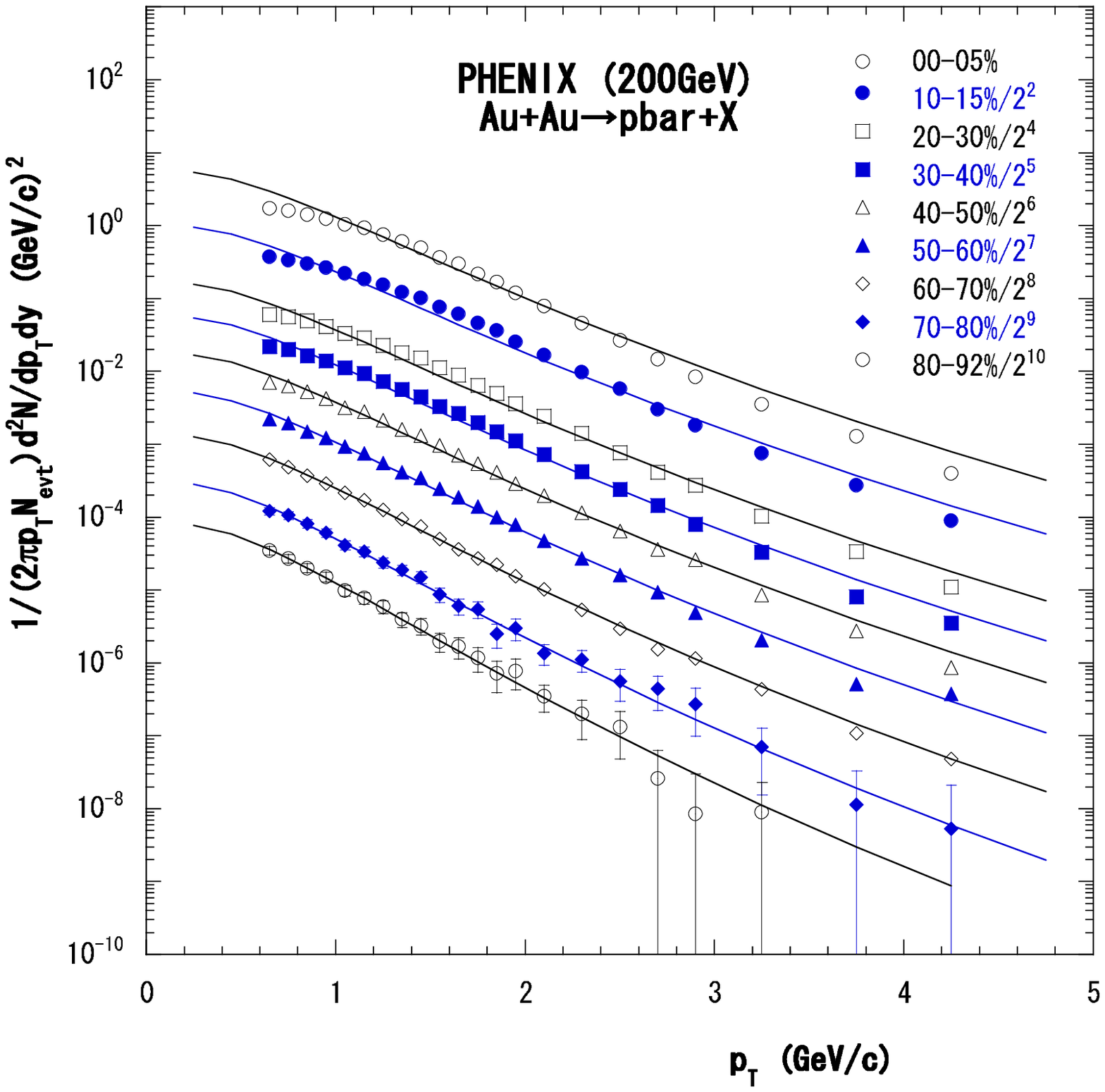} 
    \caption{\label{fig.aapbar}$p_T$ distribution 
       for $Au+Au\rightarrow\bar{p}+X$ at $y=0$~\cite{adle03c} }
 \end{figure}

The results on $Au+Au\rightarrow K^-+X$ are shown in Fig.~\ref{fig.aakn} and Table 4, and those in $Au+Au\rightarrow \bar{p}+X$ are shown in Fig.~\ref{fig.aapbar} and Table 5.  As can be seen from Figs.4 and 5, and Tables 4 and 5, fitting of the theoretical curve to the observed $p_T$ distribution are better than those in $Au+Au\rightarrow\pi^-+X$, and become much better as the centrality cut increases.
\begin{table}
 \caption{\label{tab.table3} Estimated parameters on $p_T$ distributions
  in $Au+Au\rightarrow \pi^-+X$ at $y=0$ at $\sqrt{s}=200$ GeV~\cite{adle03c}}
 \begin{ruledtabular}
 \begin{tabular}{cccc}  
  centrality  & $C$ & $\sigma(t)^2$ & $\chi^2_{min}/$n.d.f \\ \hline
  00-05\%  & 2539.6 $\pm$ 88.8 & 0.701$\pm$ 0.007 & 522.8/26\\
  05-10\%  & 2092.7 $\pm$ 73.7 & 0.710$\pm$ 0.008 & 447.4/26\\ 
  10-15\%  & 1692.0 $\pm$ 59.9 & 0.718$\pm$ 0.008 & 390.7/26\\ 
  15-20\%  & 1391.1 $\pm$ 49.9 & 0.724$\pm$ 0.008 & 319.4/26\\ 
  20-30\%  & 1052.8 $\pm$ 37.5 & 0.727$\pm$ 0.008 & 278.5/26\\ 
  30-40\%  &  667.0 $\pm$ 24.4 & 0.736$\pm$ 0.009 & 144.8/26\\ 
  40-50\%  &  409.8 $\pm$ 15.4 & 0.736$\pm$ 0.009 & 144.8/26\\ 
  50-60\%  &  241.2 $\pm$  9.3 & 0.731$\pm$ 0.010 & 114.4/26\\ 
  60-70\%  &  125.4 $\pm$  5.1 & 0.724$\pm$ 0.010 &  94.3/26\\ 
  70-80\%  &   57.4 $\pm$  2.5 & 0.719$\pm$ 0.012 &  60.4/26\\ 
  80-92\%  &   29.5 $\pm$  1.4 & 0.708$\pm$ 0.013 & 42.0/26\\ 
 \end{tabular}
 \end{ruledtabular}
 \end{table}
%

%
\begin{table}
 \caption{\label{tab.table4} Estimated parameters on $p_T$ distributions 
    in $Au+Au\rightarrow K^-+X$ at $y=0$ at $\sqrt{s}=200$ GeV~\cite{adle03c}}
 \begin{ruledtabular}
 \begin{tabular}{cccc}  
  centrality  & $C$ & $\sigma(t)^2$ & $\chi^2_{min}/$n.d.f \\ \hline
  00-05\%  & 44.75 $\pm$ 2.81 & 0.456$\pm$ 0.019 & 9.45/14\\ 
  05-10\%  & 37.05 $\pm$ 2.33 & 0.457$\pm$ 0.019 & 8.67/14\\ 
  10-15\%  & 30.82 $\pm$ 1.93 & 0.453$\pm$ 0.018 & 10.77/14\\ 
  15-20\%  & 25.69 $\pm$ 1.62 & 0.451$\pm$ 0.018 & 8.52/14\\ 
  20-30\%  & 19.17 $\pm$ 1.21 & 0.450$\pm$ 0.018 & 7.07/14\\ 
  30-40\%  & 12.92 $\pm$ 0.82 & 0.437$\pm$ 0.017 & 8.37/14\\ 
  40-50\%  &  7.86 $\pm$ 0.50 & 0.431$\pm$ 0.017 & 6.57/14\\ 
  50-60\%  &  4.55 $\pm$ 0.31 & 0.420$\pm$ 0.018 & 3.85/14\\ 
  60-70\%  &  2.27 $\pm$ 0.16 & 0.409$\pm$ 0.019 & 4.44/14\\ 
  70-80\%  &  1.01 $\pm$ 0.08 & 0.396$\pm$ 0.021 & 3.60/14\\ 
  80-92\%  &  0.50 $\pm$ 0.04 & 0.386$\pm$ 0.025 & 7.44/14\\ 
 \end{tabular}
 \end{ruledtabular}
 \end{table}
%
%
%
\begin{table}
 \caption{\label{tab.table5} Estimated parameters on $p_T$ distributions 
    in $Au+Au\rightarrow \bar{p}+X$ at $y=0$ at $\sqrt{s}=200$ GeV~\cite{adle03c}}
 \begin{ruledtabular}
 \begin{tabular}{cccc}  
  centrality  & $C$ & $\sigma(t)^2$ & $\chi^2_{min}/$n.d.f \\ \hline
  00-05\%  & 6.48 $\pm$ 0.29 & 0.296$\pm$ 0.008 & 100.1/20\\ 
  05-10\%  & 5.49 $\pm$ 0.25 & 0.296$\pm$ 0.008 & 84.31/20\\ 
  10-15\%  & 4.57 $\pm$ 0.21 & 0.297$\pm$ 0.008 & 75.89/20\\ 
  15-20\%  & 3.85 $\pm$ 0.18 & 0.296$\pm$ 0.008 & 65.83/20\\ 
  20-30\%  & 3.04 $\pm$ 0.14 & 0.287$\pm$ 0.008 & 64.19/20\\ 
  30-40\%  & 2.10 $\pm$ 0.10 & 0.281$\pm$ 0.008 & 39.75/20\\ 
  40-50\%  & 1.30 $\pm$ 0.07 & 0.277$\pm$ 0.009 & 21.33/20\\ 
  50-60\%  & 0.79 $\pm$ 0.05 & 0.266$\pm$ 0.010 & 10.40/20\\ 
  60-70\%  & 0.40 $\pm$ 0.03 & 0.254$\pm$ 0.012 &  2.23/20\\ 
  70-80\%  & 0.18 $\pm$ 0.02 & 0.239$\pm$ 0.016 &  6.29/20\\ 
  80-92\%  & 0.10 $\pm$ 0.01 & 0.225$\pm$ 0.019 &  3.18/20\\ 
 \end{tabular}
 \end{ruledtabular}
 \end{table}

In order to estimate the temperature for inclusive reactions from our analysis, we consider the approximate expression for Eq.(\ref{eq.dif5}) in the small $|{\bf p}_T|$ region.   When $\rho<<1$, $\sinh\rho\simeq\rho$. Then, Eq.(\ref{eq.dif5}) reduces to
 \begin{eqnarray}
  f(\rho,t)= C\exp\left[-\frac{\rho^2}{2\sigma(t)^2} \right].   \label{eq.dif6}
 \end{eqnarray}
Equation (\ref{eq.dif6}) should coincide with the Maxwell distribution, 
 \begin{eqnarray*}
  f({\bf v})=\left(\frac{m}{2\pi kT}\right)^{3/2}
        \exp\left[-\frac{m{\bf v}^2}{2kT} \right].   \label{eq.dif7}
 \end{eqnarray*}

Then we have an identity,
 \begin{eqnarray}
  kT=m\sigma(t)^2.   \label{eq.dif8}
 \end{eqnarray}
 \begin{figure}
    \includegraphics[scale=0.45,bb=80 160 520 650,clip]{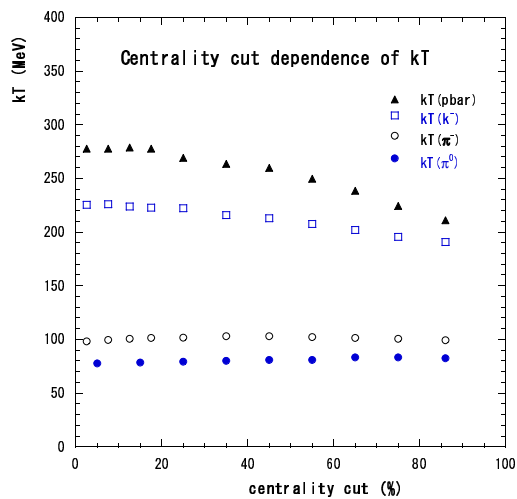} 
    \caption{\label{fig.temp} Centrality cut dependence of temperature $kT$ estimated from 
    $p_T$ distributions in $Au+Au$ collisions }
 \end{figure}

From Eq.(\ref{eq.dif8}), we can estimate the temperature $kT$ of inclusive reactions for observed particle with mass $m$.
The results are shown in Fig.~\ref{fig.temp}. We use $m_{\pi^0}=135$, $m_{\pi^-}=140$, $m_{K^-}=494$, and $m_{\bar{p}}=938$ MeV. 

The estimated temperature $kT$ from the $p_T$ distribution for $\pi^0$ is about 80 Mev, and that for $\pi^-$ is about 100 MeV. These temperatures are almost independent from the centrality cut.
On the other hand, the temperature for $K^-$ distributions gradually decreases from 230 MeV to 190 MeV, as the centrality cut increases from 5-10\% to 80-92\%.  That for ${\bar p}$ distributions decreases from 280 MeV to 210 MeV as the centrality cut increases.

Our analysis suggests that  $K^-$ and $\bar{p}$  produced at the lowest centrality cut, 0-5\%,
would keep somewhat earlier memory than $\pi^0$-meson or  $\pi^-$-meson. It is very interesting to know  whether identical particle correlations of $K$-mesons, protons and so forth depend on the centrality cut or not.
 
%
\section{Summary and discussions}

In order to analyze large $p_T$ distributions of charged particles observed at RHIC, a stochastic process in the three dimensional rapidity space is introduced. The solution is gaussian-like in radial rapidity.

Transverse momentum distributions for $\pi^0$ and $\pi^-$ at $y=0$ at $\sqrt{s}=200$ GeV observed by the PHENIX collaboration are analyzed. Observed $p_T$ distributions for $\pi^0$ in $pp$ and $Au+Au$ collisions are well described by the formula (\ref{eq.dif5}), and the result of fitting in $Au+Au$ collisions becomes much better as the centrality cut increases.

The result on transverse momentum distributions for $\pi^-$ in $Au+Au$ collisions are not good as those for $\pi^0$. However, as the centrality cut increases, the result becomes much improved.
This tendency would suggest that the initial condition (\ref{eq.dif2}) is simpler for the heavy ion collision process. At the lower centrality cut, collisions among projectile and target particles occur much more times compared with those at higher centrality cut. Therefore, secondary particles produced at each collision may have a certain (collective) velocity distribution  wider than the delta function (\ref{eq.dif5}) even in the initial stage.
In order to include the effect, we should change the initial condition (\ref{eq.dif5}), or consider the diffusion equation without radial symmetry in the three dimensional rapidity space.

\begin{acknowledgments}
 Authors would like to thank RCNP at Osaka university, Faculty of science, Shinshu university, and Matsumoto university for financial support.
\end{acknowledgments}
\end{document}